\documentclass[11pt]{article}
\pdfoutput=1
\usepackage[T1]{fontenc}
\usepackage[latin9]{inputenc}
\usepackage[a4paper]{geometry}
\usepackage[active]{srcltx}
\usepackage{amsmath}
\usepackage{amssymb}
\usepackage{esint}
\usepackage{ulem}
\usepackage{cancel}
\makeatletter

\usepackage{textcomp}

\pdfoutput=1 

\usepackage{jheppub}

\usepackage{etoolbox}
    
    \patchcmd{\maketitle}{\@fpheader}{}{}{}

\usepackage{amsfonts}

\usepackage{bbm}

\newcommand{\be}{\begin{equation}}
\newcommand{\ee}{\end{equation}}
\newcommand{\bea}{\begin{eqnarray}}
\newcommand{\eea}{\end{eqnarray}}
\newcommand\bal{\begin{aligned}}
\newcommand\eal{\end{aligned}}

\author[a,b]{Marc Henneaux,}
\author[a]{and Patricio Salgado-Rebolledo}

\affiliation[a]{Universit\'e Libre de Bruxelles and International Solvay Institutes, ULB-Campus Plaine CP231, B-1050 Brussels, Belgium}
\affiliation[b]{Coll\`ege de France, 11 place Marcelin Berthelot, 75005 Paris, France}

\emailAdd{marc.henneaux@ulb.be}
\emailAdd{psalgado@ulb.ac.be}

\preprint{}

\title{ \boldmath  Carroll contractions of Lorentz-invariant theories
   \mbox{  }    
}

\abstract
{We consider Carroll-invariant limits of Lorentz-invariant field theories. We show that just as in the case of electromagnetism, there are two inequivalent limits, one ``electric'' and the other ``magnetic''.  Each can be obtained from the corresponding Lorentz-invariant theory written in Hamiltonian form through the same ``contraction'' procedure of taking the ultrarelativistic limit $c \rightarrow 0$  where $c$ is the speed of light, but with two different consistent rescalings of the canonical variables. This procedure can be applied to general Lorentz-invariant theories ($p$-form gauge fields, higher spin free theories etc) and has the advantage of providing explicitly an action principle from which the electrically-contracted or magnetically-contracted dynamics follow (and not just the equations of motion).  Even though not manifestly so, this Hamiltonian action principle is shown to be Carroll invariant.  In the case of $p$-forms, we construct explicitly an equivalent manifestly Carroll-invariant action principle for each Carroll contraction.  While the manifestly covariant variational description of the electric contraction is rather direct,  the one for the magnetic contraction is more subtle and involves an additional pure gauge field, whose elimination modifies the Carroll transformations of the fields. We also treat gravity, which constitutes one of the main motivations of our study, and for which we provide the two different contractions in Hamiltonian form.}

\makeatother

\begin{document}
\maketitle \flushbottom

 \newpage{}

 \section{Introduction}
 The Carroll group is one of the contractions of the Lorentz group, obtained by letting the speed of light $c$ go to zero \cite{LevyLeblond:1965,Bacry:1968zf} (``ultrarelativistic limit''). It turns out to emerge in many interesting physical contexts, ranging from gravity  to condensed matter physics (see \cite{Isham:1975ur,Teitelboim:1978wv,Henneaux:1979vn} for earlier applications and 
\cite{Duval:2014uva,Grumiller:2017sjh,Ciambelli:2018wre,Bagchi:2019xfx,Donnay:2019jiz,Duval:2017els,Bagchi:2021qfe,Casalbuoni:2021fel,Pena-Benitez:2021ipo} and references therein for more recent work).

A manifestly diffeomorphism invariant formulation of a gravitation theory based on the Carroll group was given in \cite{Henneaux:1979vn}. That gravitation theory could  be viewed as the strong coupling limit  \cite{Isham:1975ur} or the ``zero signature limit'' \cite{Teitelboim:1978wv} of Einstein theory.  In that  limit, the dynamical equations obeyed by the metric involve only its time derivatives, and so, one can view this ultrarelativistic contraction as the limit of Einstein theory in which time derivatives dynamically dominate spatial gradients, a phenomenon physically relevant in the vicinity of a spacelike singularity \cite{Belinsky:1970ew,Belinsky:1982pk,Henneaux1982,Damour:2002et,Belinski:2017fas}. 

It was proved in  \cite{Duval:2014uoa} by analysing Maxwell's  equations of motion that electromagnetism in four spacetime dimensions has two inequivalent Carroll contractions, one ``electric'' and the other ``magnetic''.  This is the Carrollian analog of a similar phenomenon analysed in the Galilean limit \cite{LeBellac1973}.
The Carroll-invariant action principle was constructed in the electric case.  The Carroll transformation rules of the fields were also discussed and their difference in the two contractions was displayed.

The purpose of this note is to show that the existence of two different Carroll contractions is not peculiar to electromagnetism but that a similar phenomenon exists for all Lorentz-invariant theories whether or not they enjoy electromagnetic duality.  This phenomenon is also present  in the (full) Einstein theory of gravity, for which one can consider a ``magnetic'' contraction different from that of \cite{Isham:1975ur,Teitelboim:1978wv}, which turns out to be the ``electric'' one. 

Our method relies on the Hamiltonian formulation of the theories and on the Hamiltonian control of spacetime covariance \cite{Dirac:1962aa,Schwinger:1963xx,Teitelboim:1972vw,Bunster:2012hm}.  Our approach automatically yields in each case the relevant Carroll-invariant action principle, but being Hamiltonian, its covariance is not manifest (one cannot apply spacetime tensor calculus in the standard way). 
The problem of writing equivalent manifestly Carroll-covariant action principles for each contraction is  then solved  for $p$-form gauge fields in flat Carroll spacetime, starting with the scalar field and the electromagnetic field.   A key tool is provided by the geometrical concepts developed in \cite{Henneaux:1979vn}.   We find that the manifestly covariant description of the magnetic contractions is more subtle in that it involves an additional pure gauge field, which can be gauged away at the price of losing manifest covariance. 

Our paper is organized as follows.  After a brief survey of the geometrical concepts adapted to the description of Carroll contractions (Section \ref{sec:CarrollGeo}) and the underlying symmetry groups (Section \ref{sec:CarrollGroups}),  we establish the conditions for a theory to be Carroll-invariant (Section \ref{sec:InvHam}).
We then take the electric and magnetic Carroll contractions of various Lorentz-invariant theories, dealing successively with scalar fields, electromagnetism, $p$-form gauge fields for general $p$ (Section \ref{sec:CarrollContr}) and then higher spin gauge fields (Section \ref{sec:CarrollContrHigh}).  We find in that latter case that some of the components of the spin-$s$ fields are more conveniently put, in taking the limits, on the same footing as the conjugate momenta, i.e., should be regarded as ``$p$'s'', while their conjugate momenta should be treated as ``$q$'''s. The construction of a manifestly covariant action principle turns out to be involved especially in the magnetic case and in order to achieve this task, we focus next on $p$-form gauge fields, for which we give the covariant actions for both the electric-type and magnetic-type contractions (Section \ref{sec:manifestCov}).  We also discuss the transformations of the fields under the Carroll group, which are associated with different (dual) representations in the electric and magnetic cases. 
Section  \ref{sec:Gravity} is devoted to the explicit derivation of the Carroll magnetic limit of the Einstein theory within the Hamiltonian formalism. Finally, the last section (Section \ref{sec:Conclusions}) provides conclusions and prospects.

 \section{Carroll Geometries}
 \label{sec:CarrollGeo}
 \subsection{Zero Hamiltonian Signature Spacetimes - Minimal Carroll geometry}
Curved Carroll geometries were defined long ago in \cite{Henneaux:1979vn}.  They were called there  ``zero Hamiltonian signature spacetimes'' because the Hamiltonian signature $\epsilon = \pm 1, 0$ is a parameter that distinguishes in the Hamiltonian formulation of general relativity between Euclidean signature ($\epsilon = 1$), Lorentzian signature ($\epsilon = - 1$) and ``zero Hamiltonian signature'' ($\epsilon = 0$), which lie halfway between the Euclidean and Minkowskian cases \cite{Teitelboim:1978wv}. This is clear if one writes the spacetime metric as $g_{\alpha \beta} = \textrm{diag}(\epsilon, 1, 1, \cdots, 1)$.
 
 A vector space with a Carroll structure in $D$ dimensions involve two ingredients.
 \begin{enumerate}
 \item First, there is a degenerate metric $g_{\alpha \beta}$ of rank $D-1$ which is positive semi-definite, 
 i.e. $\det g_{\alpha \beta} = 0$, $g_{\alpha \beta} v^\alpha v^\beta \geq 0$, with $g_{\alpha \beta} v^\alpha v^\beta = 0$ if and only if the vector $v^\alpha$ is along the null direction (``null vector'').
 \item There is also a notion of normalization of the null vectors.  This normalization can be introduced in two different ways. 
 \begin{itemize}
 \item One can introduce a non-vanishing density $\Omega$ of weight one, as was done in \cite{Henneaux:1979vn}.  The ``unit'' null vector $n^\alpha$ is then such that
 \be
 \mathcal{G}^{\alpha \beta} = \Omega^2 n^\alpha n^\beta .
 \ee
 Here $\mathcal{G}^{\alpha \beta} $ are the minors of $g_{\alpha \beta}$,
 \be
 \mathcal{G}^{\alpha \beta} = \frac{1}{3!} \epsilon^{\alpha \lambda \mu \nu}\epsilon^{\beta \rho \sigma \tau} g_{\lambda \rho} g_{\mu \sigma} g_{\nu \tau}.
 \ee
 (One has clearly $\mathcal{G}^{\alpha \beta} g_{\gamma \beta} = 0$ since $\det g_{\alpha \beta} = 0$ and so the tensor density of weight two $\mathcal{G}^{\alpha \beta}$ is indeed proportional to the product $n^\alpha n^\beta$ where $n^\alpha$ is a null vector.  Giving $\Omega$ fixes the normalization of $n^\alpha$.  Note that the procedure fixes $n^\alpha$ only up to a sign.)
  \item Equivalently, if there is a time-orientation - as we shall assume from now on - , one can just give the unit (future-pointing) null vector $n^\alpha$.  
 \end{itemize}
 \end{enumerate}
 The two definitions of a Carroll structure are equivalent.  The first one explicitly shows that the number of fields characterizing a Carroll geometry ($g_{\alpha \beta}$ with $\det g_{\alpha \beta} = 0$ and $\Omega$) is equal to the number of fields characterizing a Riemannian geometry ($g_{\alpha \beta}$ with non-vanishing determinant).  The density $\Omega$ replaces the determinant of $g_{\alpha \beta}$, which is useful for writing  down variational principles.
 
 A Carroll manifold is a manifold equipped with a Carroll structure in the tangent space at each point, which depends smoothly on the point.  In local coordinates, it is defined by a symmetric tensor $g_{\alpha \beta} (x)$ with the above properties and a density $\Omega(x)$, which are both smooth.   These $\frac{D(D+1)}{2} - 1$ (components of the degenerate metric of rank $D-1$) $+1$ (volume element) $= \frac{D(D+1)}{2} $ field components match exactly in number  the components of the Riemannian metric formulation of general relativity. 

 \subsection{Raising indices and one-form $\theta_\alpha$}
 
 Because the metric is degenerate, it has no inverse, i.e., there is no tensor $g^{\alpha \beta}$ such that $g^{\alpha \beta} g_{\beta \gamma}= \delta^\alpha_\gamma$.

One can nevertheless raise indices by introducing the extra structure  of a one-form $\theta_\alpha$ such that 
\be
\theta_\alpha n^\alpha = 1
\ee
(see \cite{Henneaux:1979vn}).
One then defines the twice contravariant symmetric tensor $G^{\alpha \beta}$ such that
\be
G^{\alpha \beta} g_{\beta \gamma} = \delta^\alpha_\gamma - n^\alpha \theta_\gamma .\label{eq:DefG0}
\ee
If one imposes in addition the condition
\be
G^{\alpha \beta} \theta_\alpha \theta_\gamma = 0,
\ee
the tensor $G^{\alpha \beta}$ is completely determined.  We shall sometimes write $G^{\alpha \beta}(g_{\rho \sigma}, n^\lambda, \theta_\mu)$ to emphasize that $G^{\alpha \beta}$ depends not only on $g_{\alpha \beta}$ but also on $n^\alpha$ and $\theta_\alpha$.

One has
\be
g_{\alpha \beta} G^{\beta \gamma} v_\gamma = v_\alpha - \theta_\alpha (n^\gamma v_\gamma)
\ee
so that one gets back $v_\alpha$  after raising the index with $G^{\alpha \beta}$ and then lowering it with $g_{\alpha \beta}$ only if $v_\alpha$ is ``transverse'', i.e., $v_\alpha n^\alpha =0$.

It is useful to determine how $G^{\alpha \beta}$ changes if one changes the extra, non-Carrollian structure given by the one-form $\theta_\alpha$.  A direct computation shows that for a shift of $\theta_\alpha$,
\be \theta_\alpha \rightarrow \theta'_\alpha = \theta_\alpha + \Lambda_\alpha, \qquad \Lambda_\alpha n^\alpha = 0, \label{eq:TransfTheta}
\ee
($\Lambda_\alpha$ finite), the contravariant tensor $G^{\alpha \beta}$ transforms as
\be
G^{\alpha \beta} \rightarrow G'^{\alpha \beta} = G^{\alpha \beta} - n^\alpha \Lambda^\beta - n^\beta \Lambda^\alpha + n^\alpha n^\beta  \Lambda^\mu \Lambda_\mu, \qquad \Lambda^\alpha \equiv G^{\alpha \beta} \Lambda_{\beta} . \label{eq:TransfG}
\ee
In infinitesimal form ($\Lambda_\alpha \equiv \lambda_\alpha$ small), this becomes 
\be
\delta \theta_\alpha = \lambda_\alpha, \qquad \delta G^{\alpha \beta} =- n^\alpha \lambda^{\beta} - n^\beta \lambda^\alpha, \qquad \lambda^\beta = G^{\beta \alpha} \lambda_{\alpha}. \label{eq:TransfGInf}
\ee

The one-form $\theta_\alpha$ was actually considered more recently in the interesting work \cite{Ciambelli:2019lap}
where it was interpreted as an Ehresmann connection enabling one to split  the tangent space into the direct sum of the one-dimensional null subspace generated by $n^\alpha$ and  a unique transverse subspace to this null direction spanned by the vectors $v^\alpha$ such that $\theta_\alpha v^\alpha = 0$.  

Since the one-form $\theta_\alpha$ comes on top of the basic Carroll structure defined by the degenerate metric $g_{\alpha \beta}$ and the null vector $n^\alpha$, we shall insist that ``Carrollian physics'' should {\em not} depend on $\theta_\alpha$, i.e. should be invariant under (\ref{eq:TransfTheta}) and (\ref{eq:TransfG}). These transformations should appear as gauge transformations in any purely Carrollian action.   Using $G^{\alpha \beta}$ to raise indices might be useful in order to use tensor calculus, but one should verify in the end that the physics does not depend on which $G^{\alpha \beta}$ (i.e., which $\theta_\alpha$) is chosen.

We also recall that if the covectors $v_\alpha$ and $w_\alpha$ are both transverse, their scalar product $G^{\alpha \beta} v_\alpha w_\alpha$  does not depend on the choice of $\theta_\alpha$ and hence, is well defined in a purely Carrollian structure without need for $\theta_\alpha$.  Similarly, the trace $K_{\alpha \beta} G^{\alpha \beta}$ is independent of $\theta_\alpha$ for a transverse tensor $K_{\alpha \beta}$ ($K_{\alpha \beta}n^\alpha = 0 = K_{\alpha \beta}n^\beta$) etc.  

 \subsubsection*{Note on affine Carroll structures}
 
 In  \cite{Duval:2014uoa}, the definition of a Carroll manifold was taken to involve an additional ingredient, namely, that it should also be equipped with a symmetric affine connection preserving both the metric and the unit null vector. Since the existence of such a structure may not exist and is not unique when it exists, this brings constraints on $g_{\alpha \beta}$ \cite{Vogel1965} (see also \cite{Jankiewicz,Dautcourt1967,Dautcourt:1997hb}).  For that reason,  we shall not include that extra structure in the definition of a Carroll manifold and we shall stick to the original definition of \cite{Henneaux:1979vn} involving only the degenerate metric and a normalization of the null vectors. 

 Actually, the introduction of a metric-preserving, symmetric affine connection was also found to be unconvenient for some purposes in \cite{Duval:2014uva} and the authors of \cite{Duval:2014uoa,Duval:2014uva} reverted to the earlier definition of \cite{Henneaux:1979vn} without this extra connection, which  turns out to be appropriate for the generalization to conformal Carroll structures and the link with the BMS group \cite{Duval:2014uva}.

 \section{Carroll groups}
 \label{sec:CarrollGroups}
 
 \subsection{Infinite dimensional Carroll group $\mathcal{C}(D)$}
 
 Flat Carroll space has constant $g_{\alpha \beta}$ and $n^\alpha$.  In an appropriate coordinate system $(x^\mu)$ (``Carrollian coordinates''), one can assume
 \be
 (g_{\alpha\beta}) = \begin{pmatrix} 0 &0 \\ 0 & I_{d \times d} \end{pmatrix} , \qquad (n^\alpha) = \begin{pmatrix} 1 \\ 0 \\ 0 \\ \vdots \\ 0 \end{pmatrix} \label{eq:FlatCarroll},
   \ee
 where $ D = d+1$ and $I_{d \times d}$ is the unit matrix in $d$ dimensions.  One has then
  $\Omega = 1$.  
  
 The group $\mathcal{C}(D)$ of isometries of this structure is infinite-dimensional and given by
 \be
 x'^0 = x^0 + f(x^k), \qquad x'^k = R^k_{\; m} x^m + a^k,
 \ee
 where 
$a^k$ are constants and $ R^k_{\; m}  \in O(d)$ is an orthogonal transformation in $d$ dimensions. The function $f(x^k)$ is arbitrary. 

In infinitesimal form,
\be
\delta x^0 = \xi(x^k), \qquad \delta x^k = \omega^k_{\; m} x^m + a^k, \quad \omega_{km} = - \omega_{mk}
\ee
(with $ \omega_{km} \equiv \delta_{kt} \omega^t_{\; m}$). Note that there would be no condition on $\delta x^0$ had we only required invariance of the metric.  $\delta x^0$ could be in that case an arbitrary function of space {\em and} time. 

\subsection{Finite-dimensional Carroll group $C(D)$}
 
If one restricts the transformations to be linear, one gets the finite-dimensional (inhomogeneous) Carroll group $C(D)$, 
\be
 x'^0 = x^0 + a^0 + b_k x^k, \qquad x'^k = R^k_{\; m} x^m + a^k, \qquad a^0, a^k, b_k \in \mathbb{R}, \qquad R^k_{\; m}  \in O(d). \label{eq:CarrollTransf2}
 \ee
The parameters $b_k$ parametrize the ``Carroll boosts'', while the $a^0$ and $a^k$ are spacetime translations.  The restriction to linear transformations is natural in the flat case where the structure is defined in a vector space.  It can be implemented by requesting invariance of the flat connection $\Gamma^a_{\; bc} = 0$ which preserves both the flat metric and the flat density (and which manifestly exists in this very special flat case!).  The finite dimensional group $C(D)$ can be obtained by group contraction from the Poincar\'e group \cite{LevyLeblond:1965}. Its  homogeneous subgroupgroup $C(D) \cap GL(D)$ is obtained by setting $a^0 = a^k = 0$.

In infinitesimal form, the Carroll transformations (\ref{eq:CarrollTransf2}) read  
\be
\delta x^0 = a^0 + b_k x^k, \qquad \delta x^k = \omega^k_{\; m} x^m + a^k, \quad \omega_{km} = - \omega_{mk},
\ee
where we kept the same notation $a^0$ and $a^k$ for the infinitesimal translations and $b_k$ for the infinitesimal Carroll boosts.
 
While the flat tensors (\ref{eq:FlatCarroll}) are numerically invariant under Carroll transformations (by definition of the Carroll group), this is not so for the extra structure $\theta_\alpha$.  One finds instead that $\theta_0$ is invariant
\be
\theta'_0 = \theta_0 = 1
\ee
(in agreement with $n^\alpha \theta_\alpha = 1 = n^\alpha \theta'_\alpha$), but that the spatial components transform non-trivially,
\be
 \theta'_m = - b_k  (R^{-1})^k_{\; m}\theta_0 + \theta_k (R^{-1})^k_{\; m} = - b_k  (R^{-1})^k_{\; m}+ \theta_k (R^{-1})^k_{\; m}.
\label{eq:TransfTheta2}
\ee
Infinitesimally, one gets
\be
\delta \theta_0 = 0, \qquad \delta \theta_m = - b_m  + \theta_k \omega^k_{\; m}.
\ee
Since $\theta_0 = 1$, one can use Carroll transformations to set $\theta_k= 0$, so that $\theta_\alpha$ reads
\be
(\theta_\alpha)= \begin{pmatrix} 1 & 0 & \cdots & 0 \end{pmatrix},
\label{eq:SpecialTheta}
\ee
but this special form is not preserved in all Carroll frames (defined as frames in which $g_{\alpha \beta}$ and $n^\alpha$ take the form (\ref{eq:FlatCarroll})).  In fact, the Carroll subgroup that preserves (\ref{eq:SpecialTheta}) contains only translations and spatial rotations.

When (\ref{eq:SpecialTheta}) holds, the contravariant tensor  $G^{\alpha \beta}$ reduces to
 \be
 (G^{\alpha\beta}) = \begin{pmatrix} 0 &0 \\ 0 & I_{d \times d} \end{pmatrix}, \label{eq:FlatCarrollG}
   \ee
   but in general ($(\theta_\alpha) = (1, \theta_a)$), it reads
  \be
 (G^{\alpha\beta}) = \begin{pmatrix} \delta^{cd} \theta_c \theta_d &- \delta^{bc} \theta_c \\ - \delta^{ac} \theta_c & \delta^{ab} \end{pmatrix}, \label{eq:FlatCarrollGBis}
   \ee  
 in agreement with (\ref{eq:TransfTheta}) and (\ref{eq:TransfG}).
  It is thus not numerically invariant under Carroll boosts, for which one can actually verify that
\be
G'^{\alpha\beta}\equiv \frac{\partial x'^\alpha}{\partial x^\mu}\frac{\partial x'^\beta}{\partial x^\nu} G^{\lambda \mu} =
G^{\alpha \beta}(g_{\rho \sigma}, n^\tau, \theta'_\gamma \equiv \theta_\delta \frac{\partial x^\delta}{\partial x^\gamma}),
\ee
as it follows from (\ref{eq:CarrollTransf2}), (\ref{eq:TransfTheta2}) and (\ref{eq:FlatCarrollGBis}).

\section{Carroll  invariance in the Hamiltonian formalism}
\label{sec:InvHam}

Except when we deal with gravity, we will consider in this paper Carroll invariant dynamics of fields $\phi^A(x)$  in flat Carroll spacetime. When these are tensor fields, their transformation under a linear Carroll transformation,
\be
x'^\beta = C^\beta_{\; \alpha} x^\alpha + a^\beta, \qquad \left(C^\beta_{\; \alpha}\right) = \begin{pmatrix} 1 & b_a \\ 0 & R \end{pmatrix},
\ee
is just inherited from their transformation under general coordinate transformations.  For instance, we saw that for a one-form $\theta_\alpha$,
\be
\theta'_\alpha (x') =(C^{-1})^\beta_{\; \alpha} \theta_\beta(x), \Leftrightarrow \theta'_0(x') = \theta_0(x) , \quad \theta'_a (x') = (R^{-1})^b_{\; a}\left(- b_b \theta_0+ \theta_b(x) \right),
\ee
which is just the restriction to a linear Carroll transformation of the general coordinate transformation
\be
\theta'_\alpha (x') = \frac{\partial x'\beta}{\partial x^\alpha} \theta_\beta(x).
\ee
For a vector field, one has
\be
V'^\alpha (x') =C^\alpha_{\; \beta} V^\beta(x), \Leftrightarrow V'^0(x') = V^0(x) + b_a V^a(x) , \quad V'^a (x') = R^a_{\; b} V^b(x) ,
\ee
and the product $\theta_\alpha V^\alpha$ is clearly invariant. 
Under an infinitesimal Carroll transformation parametrized by the vector field $\xi^\mu$, the fields $\phi^A$ transform with the Lie derivatives,
\be
\delta_\xi \phi^A = \mathcal{L}_\xi \phi^A.
\ee

A Carroll transformation is generated in the canonical formalism by 
\be
a^0 E + a^k P_k + b_k B^k + \frac12 \omega_{k m} M^{km},
\ee
where the Carroll generators are given by integrals of local densities involving the ``energy density'' $\mathcal{E} (x)$ and the ``momentum density'' $\mathcal{P}_k (x)$. The spacetime translations are generated by 
\be
E = \int d^d x \mathcal{E} (x) , \qquad P_k = \int d^d x \mathcal{P}_k (x), 
\ee
while the generators of Carroll boosts and spatial rotations read
\be
 B^k =  \int d^d x x^k \mathcal{E} (x) , \qquad M^{rs} = \int d^d x (x^r \delta^{sk} - x^s \delta^{rk})\mathcal{P}_k (x)
\ee
In particular, the dynamical generator $E$ is the generator of time translations (Hamiltonian) and depends on the action.  

A necessary and sufficient condition for the theory to be Carroll invariant is that the generators fulfil the Carroll algebra
\begin{eqnarray}
&& [P_k, B^m] = \delta^m_{\; k} E, \\
&&  [P_k, M^{rs}]= (\delta^r_k \delta^{sl} -\delta^s_k \delta^{rl}) P_l, \qquad [B^k, M^{rs}]=  - B^r \delta^{sk} + B^s \delta^{rk},\qquad \\
&& [M^{km}, M^{rs}] = - \delta^{kr} M^{ms} +\delta^{mr} M^{ks} +\delta^{ks} M^{mr} -\delta^{ms} M^{kr}
\end{eqnarray}
(other Poisson brackets equal to zero). Indeed, when this is the case, the Hamiltonian action $S[\phi^A, \pi_A] =
\int dx^0 (\int d^dx\,  \pi_A \dot{\phi}^A - E)$ is invariant under the canonical transformations generated by $E, P_k, B^m, M^{rs}$ and these transformations close in the same way as their generators, i.e., according to the Carroll algebra. Here $\pi_A$ is the momentum conjugate to $\phi^A$.

The condition that the generators $E, P_k, B^m, M^{rs}$ should close according to the Carroll algebra implies constraints on the form of the Poisson brackets of the densities $\mathcal{E} (x)$ and  $\mathcal{P}_k (x)$ out of which they are constructed, just as in the Lorentz invariant case \cite{Dirac:1962aa,Schwinger:1963xx}.  

In fact, since $P_k = \int d^d x\mathcal{P}_k (x)$ is a kinematical generator, the form of which can be determined without knowing the action, the non trivial conditions for Carroll covariance are only conditions on $\mathcal{E}$ and are fulfilled if (i) $\mathcal{E}$ is a scalar under spatial translations and rotations (kinematical transformations); and (ii)
\be
[\mathcal{E} (x), \mathcal{E} (x')] = 0
\ee
(to be compared with the Dirac-Schwinger conditions $[\mathcal{E} (x), \mathcal{E} (x')] \sim ( \mathcal{P}^k(x) +  \mathcal{P}^k(x')) \delta_{,k}(x-x')$). 
We now establish that these two conditions imply indeed the Carroll algebra.

The kinematical momentum density $\mathcal{P}_k (x)$ is given by $\int d^d x \xi^k \mathcal{P}_k  = \int d^3 x \pi_A {\mathcal L}_\xi \phi^A$, so that $[\phi^A(x), \int d^d y\xi^k(y) \mathcal{P}_k (y)] = {\mathcal L}_\xi \phi^A(x)$ and $[\pi_A(x), \int d^d y\xi^k(y) \mathcal{P}_k (y)] = {\mathcal L}_\xi \pi_A(x)$ for any spatial vector $\xi^k(x)$.   This implies quite generally $[F(x), \int d^d y\xi^k(y) \mathcal{P}_k (y)] = {\mathcal L}_\xi F(x)$ for any function of the fields.  Since $\int d^d y\xi^k(y) \mathcal{P}_k (y)$ is the generator of the spatial Lie derivatives, the algebra of the kinematical generators $P_k$ and $M^{rs}$ is automatically fulfilled if we specialize $\xi^k$ to be a spatial translation or rotation.

Now, if $F$ is a scalar under spatial translations or rotations, one gets
\be
[F(x), \int d^d y(a^k \mathcal{P}_k (y)+ \omega_{r}^{\; k} x^r\mathcal{P}_k (y)] =(a^k  + \omega_{r}^{\; k} x^r) \partial_k F(x),
\ee
i.e.,
\be
[F(x), a^k  P_k + \frac12 \omega_{rs} M^{rs} ] =(a^k  + \omega_{r}^{\; k} x^r) \partial_k F(x).
\ee
This equation holds in particular  for $\mathcal{E}$ if it is a scalar, as condition (i) above imposes.  Integrating then this Poisson bracket  over  space  gives $\int d^dx [\mathcal{E}, a^k  P_k + \frac12 \omega_{rs} M^{rs} ] =\int d^d x(a^k  + \omega_{r}^{\; k} x^r) \partial_k \mathcal{E}$ and an integration by part of the right hand side yields  the correct Poisson brackets $[E, P_k] $  and $[E, M^{rs} ]$ (which both vanish).  Similarly, multiplying $\mathcal{E}$ by $x^m$, integrating over space, using the above Poisson bracket and integrating by parts yields also the correct Poisson brackets $[B^m, P_k]$ and  $[B^m, M_{rs}]$.  [We assume the energy density to decay sufficiently fast so that the integrals for the Carroll generators converge and the surface integrals occurring  at infinity in this computation vanish.] 

The second condition (ii)  implies even more straightforwardly that the Poisson brackets $[E, B_k] $ and $[B_k, B_m] $ are equal to zero.  This shows that the full Carroll algebra is satisfied.

We note that in the case of a gauge invariant theory, a further consistency condition should be verified, namely, that the Carroll generators should be gauge invariant, i.e., have vanishing Poisson brackets with the generators of gauge transformations up to these generators themselves (which weakly vanish), leaving thereby invariant the corresponding constraint term in the action (with possibly a transformation of the Lagrange multipliers).  This will be the case if the integrands  of the Carroll generators themselves are  gauge invariant, or gauge invariant up to a total derivative (modulo the gauge constraints).

We now describe the two different types of contractions of Lorentz-invariant field theories.

 \section{Carroll contractions of $p$-form gauge theories}
 \label{sec:CarrollContr}
 
\subsection{Scalar field}
\label{sec:Scalar}

We start with the Klein-Gordon field.  Since there is no universal speed to convert time into length in the limit $c \rightarrow 0$, we use a time variable $t$ that has dimension of time and keep track explicitly of the powers of $c$ in the Lorentz-invariant action before taking the Carrollian limits.  The Lorentzian metric reads $ds^2 = - c^2 dt^2 + \sum_k (dx^k)^2$ (in particular, $\eta_{tt} = - c^2$)
and the inverse component $\eta^{tt}$ is equal to $-\frac{1}{c^2}$. We initially assign a dimension to the scalar field such that the potential energy density $(\partial_k \phi)^2$ has units of an energy density (without power of $c$).  This choice is adapted to the magnetic-type Carroll contraction. A different choice will be made below when we take the electric-type contraction.
 
The canonical action for a scalar field in Minkowski space is,
\be
S[\phi, \pi_\phi] = \int dt \left[\int d^d x \pi_\phi \dot{\phi} - H\right], \qquad \dot{\phi} \equiv \partial_t \phi,
\ee
with
\be
H = \int d^d x \mathcal{E}, \qquad \mathcal{E} = \frac12 \big[ c^2\left( \pi_\phi \right)^2 + \partial_k \phi \partial^k \phi  \big],
\ee
where the indices are raised with the flat metric.

The magnetic contraction is the straightforward limit $c \rightarrow 0$ in that expression, which gives
\be
S^M[\phi, \pi_\phi] = \int dt \left[  \int d^d x \pi_\phi \dot{\phi} - H^M\right]  \label{eq:MagnAction0},
\ee
\be
H^M = \int d^d x \mathcal{E}^M, \qquad \mathcal{E}^M = \frac12  \partial_k \phi \partial^k \phi .
\ee
The limit is direct because one only sets to zero the ``visible'' $c$'s, without rescaling  the field $\phi$ or its conjugate momentum $\pi$.

Alternatively, by rescaling $\phi = c \phi'$, $\pi_\phi = \frac{1}{c} \pi'_\phi$, which preserves the canonical structure,  taking the limit $c \rightarrow 0$ and dropping then the primes, one gets the electric contraction
\be
S^E[\phi, \pi_\phi] = \int dt \left[\int d^d x \pi_\phi \dot{\phi} - H^E\right],
\ee
\be
H^E = \int d^d x \mathcal{E}^E, \qquad \mathcal{E}^E = \frac12  \left( \pi_\phi \right)^2  .
\ee
[In terms of the old variables $\mathcal{E}^E$ is equal to $\frac12  \left( c \pi_\phi \right)^2$; the change of variables absorbs the factor $c^2$ and ``transfers'' it to $\mathcal{E}^M$, which reads $ \frac{c^2}{2}  \partial_k \phi \partial^k \phi$ in terms of the new variables.]   

The terminology ``electric contraction'' and ``magnetic contraction'' is used in analogy with the terminology introduced in \cite{Duval:2014uoa} for electromagnetism.  The field $\phi$ has naturally different units in the electric and magnetic contractions, since it is either $\dot{\phi}^2$ or $(\partial_k \phi)^2$ that has the dimension of an energy density.

Both contractions are Carroll-invariant since the resulting energy density $\mathcal{E}(x^k)$ obeys in both cases the Poisson brackets
\be
[\mathcal{E}(x^k), \mathcal{E}(x'^k)]= 0,
\ee
characteristic of Carroll-invariant dynamics.  The momentum density $\mathcal{P}_k$ generating spatial Lie derivatives is given by
\be
\mathcal{P}_k = \pi_\phi \partial_k \phi.
\ee

Both limits are also compatible with Carroll causality, which requests that information propagates only along the null curves (neighbouring points do not speak to each other).
Indeed, in the magnetic case, the equations are $\dot{\phi} = 0$ and $\dot{\pi}_\phi =  \triangle \phi$.  This implies $\phi(t, x^k) = \phi(0, x^k)$ and $\pi_\phi (t, x^k) =  t \triangle \phi (0, x^k) + \pi_\phi (0, x^k)$, which shows that the fields at time $t$ and space $x^k$ depend only on the fields (and a finite number of their spatial derivatives) at time $t=0$ evaluated at the same spatial point $x^k$. 
In the electric case, the equations of motion are $\dot{\phi} = \pi_\phi$, $\dot{\pi}_\phi = 0$ and lead to similar conclusions\footnote{The wave equation obtained after integrating out $\pi_\phi$ in the electric case has been considered in \cite{Bergshoeff:2014jla}.}.

There are thus two ways to produce a Carroll-invariant theory.  Either we drop the spatial gradients of the fields in the energy density and keep only the time derivatives, i.e. the conjugate momenta (``electric limit'').  Or we drop the conjugate momenta and keep only the spatial gradients (``magnetic limit'').  The kinetic term $\int dt \int d^d x \pi_\phi \dot{\phi} $ in the action is always kept intact.  In both cases the key relation $[\mathcal{E}(x^k), \mathcal{E}(x'^k)]= 0$ is obviously fulfilled since the resulting energy density `depends only on the ``$p$'s'' or on the ``$q$'s''.  Furthermore, since both the kinetic energy density and the potential energy density are independently scalars under spatial translations and rotations, the whole Carroll algebra is fulfilled.

Note that a mass term $m^2 \phi^2$ is acceptable in both contractions provided one rescales the mass so that this term survives in the limits.  This is true even in the electric-type Carrollian limit because the resulting energy density contains undifferentiated $\pi$ and $\phi$ (it is ``ultralocal'').  The mass must be rescaled with different powers of $c$ in the magnetic and electric  limits, in such a way that $m^2 \phi^2$ has the dimensions of an energy density in terms of the original $\phi$ (magnetic contraction) or the new one (electric contraction).  
 
 \subsection{Electromagnetism}
  \label{sec:EM}
 
We start with the standard Lorentz-invariant Maxwell action in Hamiltonian form, which reads
 \be
 S[A_i, \pi^i, A_t ] = \int dt\left[ \int d^d x \pi^a \dot{A}_a -H\right],
 \ee
 where
 \be
 H=\int d^d x (\mathcal E- A_t \partial_a \pi^a) , \qquad \mathcal E= \frac12 \left(c^2\pi^a \pi_a + \frac12 F^{ab} F_{ab} \right) 
 \ee
($A_t dt$ and $A_a dx^a$ have same dimension, i.e., $[A_t] = [c A_a]$, so that $\partial_i A_t$ and $ \dot{A}_a$ have same dimension). Here $\pi^a$ is the momentum conjugate to $A_a$ and is (up to the factor $c^{-2}$ and on the Maxwell shell) equal to the mixed temporal-spatial components of the field strength $F_{ta}$,
\be
F_{ta} = \dot{A}_a -\partial_a A_t,
\ee
whereas $F_{ab}$ is the magnetic field defined as usual,
\be
F_{ab} = \partial_a A_b - \partial_b A_a.
\ee
 
 The magnetic contraction is obtained by letting $c \rightarrow 0$, which yields
  \be
 S^M[A_a, \pi^a, A_0] = \int dt\left[ \int d^d x \pi^a \dot{A}_a -H^M\right] ,
 \label{eq:MagneticAction00}
 \ee
  where
 \be
 H^M=\int d^d x (\mathcal E^M- A_t \partial_a \pi^a) , \qquad \mathcal E^M= \frac14 F^{ab} F_{ab} ,
  \ee
  so that
   \be
[\mathcal{E}^M(x^k), \mathcal{E}^M(x'^k)]= 0.
\ee
The field equations that follows by extremizing the Carroll magnetic action are then given by
\be
\bal
&\delta \pi^i : \quad F_{ta} \equiv \dot{A}_a -\partial_a A_t =0,\\
&\delta A_a:\quad \dot \pi^a-\Delta A^a + \partial^a\partial_b A^b=0, \\
&\delta A_t:\quad \partial_a \pi^a=0.
\eal
\ee
In that limit, $\pi^a$ is no longer equal to $F_{ta}$. We still call it, however, the ``electric field''  (or rather, minus the electric field), since $F_{ta}$, being zero, is not a particularly interesting object.

In four spacetime dimensions, one can equivalently rewrite these equations of motion in terms of the electric field $E^a = - \pi^a$ and the magnetic field $B^a = \frac12 \epsilon^{abc}F_{bc} = \epsilon^{abc} \partial_b A_c$ as 
 \be
\boldsymbol{\nabla} \cdot \boldsymbol{ E} = 0, \qquad \boldsymbol{\nabla} \cdot \boldsymbol{ B} = 0,  \qquad \frac{\partial \boldsymbol{ E} }{\partial t} -\boldsymbol{\nabla} \times \boldsymbol{ B} = 0, \qquad \frac{\partial \boldsymbol{ B} }{\partial t} = 0,
\ee
in agreement with \cite{Duval:2014uoa}. (The last equation follows from $F_{ta} = 0$ and the Bianchi identity;  conversely it implies $F_{ta} = 0$ by a suitable choice of $A_t$.)  
 
 To reach the electric-like contraction, we rescale the fields as $A_a\rightarrow c A_a$, $\pi^a\rightarrow \frac{1}{c}\pi^a$ and also $A_t\rightarrow c A_t$. In the limit $c\rightarrow 0$ this yields 
  \be
 S^E[A_a, \pi^a, A_t] = \int dt \left[ \int d^d x \pi^a \dot{A}_a -H^E\right],
 \ee
  where
 \be
 H^E=\int d^d x (\mathcal E^E- A_t \partial_a \pi^a) , \qquad \mathcal E^E= \frac12 \pi^a \pi_a ,
 \ee
 so that
   \be
[\mathcal{E}^E(x^k), \mathcal{E}^E(x'^k)]= 0.
\ee
 The field equations are now
 \be
\bal
&\delta \pi^a : \quad \dot A_a -\partial_a A_t-\pi_a =0,\\
&\delta A_a:\quad \dot \pi^a=0,\\
&\delta A_0:\quad \partial_a \pi^a=0,
\eal
\ee
and, in four spacetime dimensions, are equivalent to
 \be
\boldsymbol{\nabla} \cdot \boldsymbol{ E} = 0, \qquad \boldsymbol{\nabla} \cdot \boldsymbol{ B} = 0,  \qquad \frac{\partial \boldsymbol{ B} }{\partial t} +\boldsymbol{\nabla} \times \boldsymbol{ E} = 0, \qquad \frac{\partial \boldsymbol{ E} }{\partial t} = 0,
\ee
again in agreement with \cite{Duval:2014uoa}.

Since the energy density $\mathcal{E}(x)$ fulfills in both cases the Poisson bracket relation $[\mathcal{E}(x), \mathcal{E}(x')] = 0$ and is gauge invariant, it can be used to construct,  together with the momentum density $\mathcal{P}_k = F_{km} \pi^m$, the Carroll generators $E, P_a, B^a, M^{rs}$.   Note that this choice of $\mathcal{P}_k$, which is gauge invariant, differs from the generator of spatial diffeomorphisms by a physically irrelevant term proportional to the gauge constraint-generator $-\partial_a \pi^a \approx 0$ (Gauss law).  The standard generator of spatial diffeomorphisms could be equally used, a kinematical issue that is actually independent of whether one performs or not  a Carroll contraction.

Carroll causality  is meaningful only for gauge-invariant functions (observables).  It is discussed as in the scalar field case by integrating explicitly the field equations.  Given the similarity between the equations of motion of the electric and magnetic Carroll contractions, it is sufficient to consider only one of them, say the electric limit.  In that case, one gets
\be
B^a(t,x^k)=t \epsilon^{abc}\partial_b\pi_c(0,x^k)+B^a(0,x^k) , \qquad \pi^a(t,x^k)=\pi^a(0,x^k),
\ee
the compatibility with Carroll invariance being then manifest since information evidently propagates only along the lines $x^k = const.$

One can easily include a mass term.  The Proca Lagrangian is the above Lagrangian supplemented by $\frac12 m^2 (c^{-2}A_t^2 -   A_a A^a)$ (in appropriate units for the mass).  There is no gauge invariance and the field $A_0$ is an auxiliary field that can be eliminated using its own equation of motion, to give
\be
S^{\textrm{Proca}}=\int dt \, d^d x \left(\pi^a \dot A_a-\mathcal H^{\textrm{Proca}} \right),
\ee  
with
\be
\mathcal H^{\textrm{Proca}} =  \mathcal E^E + \mathcal E^M,
\ee
and
\be
 \mathcal E^E =  c^2 \left(\frac{1}{2}\pi_a \pi^a +  \frac{1}{2m^2} (\partial_a\pi^a)^2\right),   \qquad \mathcal E^M = \frac{1}{4}F_{ab}F^{ab} +\frac{m^2}{2}A_a A^a  .
\ee
One clearly has $[\mathcal{E}^C(x^k), \mathcal{E}^C(x'^k)]= 0$ for $C = M$ or $E$. Both magnetic-type and electric-type Carroll limits can be taken as above (no rescaling for the magnetic case; rescaling $\pi^i \rightarrow c^{-1} \pi^i$, $A_i \rightarrow c A_i$ prior to taking the limit for the electric case, in order to pass the $c^2$-factor from $\mathcal E^E$  to $\mathcal E^M$ in terms of the rescaled variables). 

\subsection{$p$-form gauge fields (general $p$) and interactions}

We shall use from now on the notation $x^0 \equiv t$ (without factor of $c$) in order to avoid possible confusion between $\partial_t = \frac{\partial}{\partial t}$ ($ t$ time) and $\partial_t = \frac{\partial}{\partial x^t}$ ($t$ latin index).  We also write $d^D x \equiv dt d^{d}x $.

\subsubsection*{$p$-form electrodynamics}
We consider a $p$-form gauge field
\be
A=\frac{1}{p!}A_{\alpha_1\cdots\alpha_p} 
dx^{\alpha_1}\wedge\cdots\wedge dx^{\alpha_p},
\ee
whose associated field strength is given by the curvature $(p+1)$-form
\be
F=dA= \frac{1}{(p+1)!}F_{\alpha_1\cdots\alpha_{p+1}}
dx^{\alpha_1}\wedge\cdots\wedge dx^{\alpha_{p+1}},
\ee
where
\be
F_{\alpha_1\cdots\alpha_{p+1}}=(p+1)\partial_{[\alpha_1}
A_{\alpha_2\cdots\alpha_{p+1}]}.
\ee

The action for the free theory,
\be\label{Spform}
\bal
S[A_{\alpha_1\cdots\alpha_p}]&=-\frac{1}{2(p+1)!}\int d^D x F_{\alpha_1\cdots \alpha_{p+1}}  F^{\alpha_1\cdots \alpha_{p+1}}
\\&
=\frac{1}{2}\int d^D x \left( 
\frac{1}{c^2p!}F_{0a_1\cdots a_p}  F_0^{\;\; a_1\cdots a_p}
-\frac{1}{(p+1)!}F_{a_1\cdots a_{p+1}}  F^{a_1\cdots a_{p+1}}
\right),
\eal
\ee
can be cast in Hamiltonian form following the standard procedure.  One finds
\be
\bal
S[A_{a_1\cdots a_p}, \pi^{a_1\cdots a_p},A_{0a_1\cdots a_{p-1}}] &=\int d^D x 
\left(
\pi^{a_1\cdots a_p} \dot A_{a_1\cdots a_p}
-  A_{0a_2\cdots a_{p}}\mathcal{G}^{a_2 \cdots a_{p}} -\mathcal H
\right),
\eal
\ee
where
\be
\mathcal H = 
 \mathcal{E}^E
+\mathcal{E}^M,
\ee
with
\be
\mathcal{E}^E = \frac{p!c^2}{2}\pi_{a_1\cdots a_p}  \pi^{ a_1\cdots a_p}, \quad \mathcal{E}^M = \frac{1}{2(p+1)!}F_{a_1\cdots a_{p+1}}  F^{a_1\cdots a_{p+1}} ,
\ee
and 
\be
\mathcal{G}^{a_1 \cdots a_{p-1}} = - p \partial_{a} \pi^{a a_1 \cdots a_{p-1}}.
\ee
The variables $\pi^{a_1\cdots a_p}$ are the momenta conjugate to $A_{a_1\cdots a_p}$ while $A_{0a_1\cdots a_{p-1}}$ are the Lagrange multipliers for the constraints $\mathcal{G}^{a_1 \cdots a_{p-1}}  \approx 0$.

The magnetic Carrollian limit is straightforward, whereas the electric limit requires the rescalings
\be
A_{a_1\cdots a_p} \rightarrow c A_{a_1\cdots a_p}  ,  \quad
\pi^{a_1\cdots a_p} \rightarrow \frac{1}{c} \pi^{a_1\cdots a_p}  ,  \quad
A_{0a_1\cdots a_{p-1}} \rightarrow c A_{0a_1\cdots a_{p-1}} .
\ee

\subsubsection*{Interactions}

Carrollian contractions are also compatible with the switching on of interactions (when these consistently exist).  We explicitly consider here the Yang-Mills case, and $p$-form interactions.  

\begin{itemize}
\item {\bf Yang-Mills field:}
The Yang-Mills action for a non-Abelian gauge field $A_{\alpha}=A_\alpha^A T_A$, where $T_A$ stands for some set of Lie algebra generators, 
$[T_A, T_B]= f^C_{\;\;AB} T_C$,  is given by
\be\label{actionYM}
S=-\frac{1}{4g^2}\int d^D x F^A_{\alpha\beta}F_A^{\alpha\beta}  , \qquad F^A_{\alpha\beta}=\partial_\alpha A^A_\beta -\partial_\beta A^A_\alpha+ [A_\alpha, A_\beta]^A ,
\ee
where $g$ is the Yang-Mills coupling constant. For comparison with the abelian case, we perform the convenient rescaling $A^A_\alpha \rightarrow g A^A_\alpha$, $F^A_{\alpha\beta} \rightarrow g F^A_{\alpha\beta}$, which yields
\be\label{actionYMrescg}
S=-\frac{1}{4}\int d^D x F^A_{\alpha\beta}F_A^{\alpha\beta}  , \qquad F^A_{\alpha\beta}=\partial_\alpha A^A_\beta -\partial_\beta A^A_\alpha+g [A_\alpha, A_\beta]^A.
\ee

The Hamiltonian action then takes the form
\be\label{SYMresc1}
S[A^A_a, \pi^a_A, A^A_0] =\int d^D x \left(\pi^a_A \dot A^A_a-\mathcal H -A^A_0 \mathcal{G}_A \right) , \quad \mathcal H =\frac{c^2}{2}\pi^A_a \pi_A^a+ \frac{1}{4}F^A_{ab}F_A^{ab} ,
\ee
where $\pi^a_A$ are the momenta conjugate to $A^A_a$ and $A^A_0$ the Lagrange multiplier for the non-abelian Gauss constraint $\mathcal{G}_A \approx 0$, with 
\be
\mathcal{G}_A = -  D _a \pi^a_A \equiv - (\partial_a \pi^a_A +g\,f^C_{\;\;BA} A^B_a \pi^a_C).
\ee
The constraints generate the Yang-Mills gauge transformations. 

The magnetic Carrollian contraction is straightforward in \eqref{SYMresc1}.  The electric contraction requires to set 
\be\label{rescElnonab}
A^A_a\rightarrow c A^A_a , \qquad \pi^a_A\rightarrow \frac{1}{c} \pi^a_A , \qquad A^A_0\rightarrow c A^A_0.
\ee
However this renders the curvature $F_{ab}^A$, as well as the covariant derivative $D_a$ Abelian. While this result yields a consistent theory, one can circumvent it by supplementing \eqref{rescElnonab} with
\be
g\rightarrow \frac{1}{c} g.
\ee
The effect of this rescaling is that the constraint-generator $\mathcal{G}_A$ does not rescale, whereas $F_{ab}^A$ rescales in the same way as $A_a^A$. The action then takes the form
\be
\bal
S&=\int d^D x \left(\pi^a_A \dot A^A_a-\mathcal H +A^A_0 D _a \pi^a_A \right) , \qquad \mathcal H &=\left(\frac{1}{2}\pi^A_a \pi_A^a+ \frac{c^2}{4}F^A_{ab}F_A^{ab} \right),
\eal
\ee
on which the electric Carrollian contraction can be implemented. 

Note that one can introduce a mass in term in the Yang-Mills action in complete analogy to the Abelian case.

\item {\bf $p$-form interactions:} We consider for definiteness the coupled Yang-Mills-$2$-form system, with action
\be
S[A^A_\alpha, B_{\alpha\beta}] =\int d^D x \left(
-\frac{1}{4} F^A_{\alpha\beta} F_A^{\alpha\beta}
-\frac{1}{12} \left(G_{\alpha\beta\gamma} 
+\lambda \Theta_{\alpha\beta\gamma}\right)\left(G^{\alpha\beta\gamma}
+\lambda \Theta^{\alpha\beta\gamma}\right)
\right).
\ee
 Here, $A^A_\alpha$ is the Yang-Mills field  and $B_{\alpha\beta}$ is the Abelian two-form, $F^A_{\alpha\beta}$ is the Yang-Mills curvature tensor given in \eqref{actionYMrescg}, $G_{\alpha\beta\gamma}$ is the field strength of the $2$-form,
\be
G_{\alpha\beta\gamma}= \partial_\alpha B_{\beta\gamma}
- \partial_\beta B_{\alpha\gamma}- \partial_\gamma B_{\beta\alpha} , 
\ee 
$\lambda$ is a constant and $\Theta_{\alpha\beta}$ are the components of the Chern-Simons form
\be 
\bal
\Theta&= \frac{1}{3}{\rm Tr}\left[A\wedge F-\frac{1}{6} A\wedge [A,A]\right],
\eal
\ee
where, after implemented the rescaling $A^A_\alpha \rightarrow g A^A_\alpha$, $F^A_{\alpha\beta} \rightarrow g F^A_{\alpha\beta}$ as in \eqref{actionYMrescg} and absorbing a factor $g^2$ in $\lambda$, we can write
\be
\Theta_{\alpha\beta\gamma}=
A_{A[\alpha}F^A_{\beta\gamma]} -\frac{g}{3} f_{ABC} A^A_{\alpha}A^B_{\beta}A^C_{\gamma}.
\ee
We have also assumed the gauge group to be compact so that $f_{ABC}$ is totally antisymmetric. 

The Hamiltonian form of the action has been worked out in \cite{Baulieu:1986hp} and reads in obvious notations
\be
\bal
S_H=\int d^D x& \bigg(
\pi^{a}_A\dot A^A_a +P^{ab}\dot B_{ab} -\mathcal H
-2B_{0a}\partial_b P^{ab}+A^A_{0}\left(D_a \pi^{a}_A
+2\lambda P^{ab} \partial_a A_{Ab} \right)
\bigg),
\eal
\ee
where the Hamiltonian density is given by 
\be
\bal
\mathcal H &=  \mathcal E^E +  \mathcal E^M \\
\mathcal E^E &=  c^2 \left(P_{ab} P^{ab}
 + \frac{1}{2} (\pi^{a}_A-2\lambda P^{ab} A_{Ab})
 (\pi^A_a- 2\lambda  P_{ac} A^{Ac}) \right)\\
 \mathcal E^M &=\frac{1}{4}F^A_{ab}F_A^{ab}+\frac{1}{12} \left(G_{abc} 
+\lambda \Theta_{abc}\right)\left(G^{abc}
+\lambda \Theta^{abc} \right).
\eal
\ee

One has 
\be
[\mathcal E^M (x^k), \mathcal E^M (y^k)] = 0, \qquad [\mathcal E^E (x^k), \mathcal E^E (y^k)] = 0.
\ee
In the electric case, where both $A_a^A$ and its conjugate momentum appear, this commutation relation holds because no derivative of the canonical variable enter $\mathcal E^E$ (ultralocality).   

We can immediately see that the magnetic Carrollian limit is again straightforward, whereas the electric one requires the rescalings
\be
\bal
&A^A_a\rightarrow c A^A_a , \qquad \pi_A^a\rightarrow \frac{1}{c} \pi_A^a,\\&
B_{ab}\rightarrow c B_{ab} , \qquad P^{ab}\rightarrow \frac{1}{c} P^{ab},\\&
A^A_0\rightarrow c A^A_0 , \qquad
B_{0a}\rightarrow c B_{0a} , \\&
\lambda \rightarrow \frac{1}{c} \lambda
 , \qquad
g \rightarrow \frac{1}{c} g
\eal
\ee
(the rescalings of the coupling constants guarantee that the interactions survive in the limit).

We considered here the explicit case of the Chern-Simons-like couplings of a $1$-form with a $2$-form.  The analysis can readily be extended to more general form degrees.

\end{itemize}

\section{Carroll contractions of  higher spin gauge theories}
\label{sec:CarrollContrHigh}

The reasons why the above contraction procedure can be applied without difficulty, yielding consistent Carroll-invariant theories with a variational description, can be characterized as follows.

The Hamiltonian formulation of these Lorentz-invariant theories involves an energy density and a momentum density.  While the momentum density is unaffected in the Carroll contractions, this is not so for the energy density.

The energy density is a sum of two terms, each of which is a scalar under the kinematical spatial translations and rotations.  These are the potential energy density containing the fields and the kinetic energy density containing their conjugate momenta. The potential energy densities at distinct spatial points have vanishing Poisson brackets, as do the kinetic energy densities - but the brackets between the potential and kinetic energy densities do have non-trivial Poisson bracket that ensure the validity of the Dirac-Schwinger Poisson bracket relations.
Therefore, if one drops either the potential energy density (electric contraction) or the kinetic energy density (magnetic contraction), one gets a Carroll-invariant theory.  

The full consistency of the procedure is established once one has verified that it is compatible with gauge invariance.  Now, the gauge generators are unchanged in the Carroll limits and the gauge transformations in phase space remain the same.   In the verification that the total (kinetic + potential) energy density yields gauge invariant generators $\int d^d x \mathcal{E} (x)$ (Lorentz energy) and $\int d^d x x^a \mathcal{E} (x))$ (Lorentz boost generator at $x^0 =0$), which does hold because of the consistency of the pre-contraction Lorentz-invariant theory,  there is no compensation between the individual contribution of each type of energy density, because the gauge transformations of the fields and their momenta involve independent parameters.  Therefore, each type of energy densities leads to gauge invariant Carroll generators.

These properties also hold for higher spin gauge fields described by the relativistic Fronsdal action  \cite{Fronsdal:1978rb} and therefore these theories also possess two different consistent Carroll limits, one electric and one magnetic. 

We illustrate the contraction procedure in the Pauli-Fierz case (spin $2$) and in the spin $3$ case.

\subsection{Pauli-Fierz field}

Keeping track of the powers of $c$, one finds that the massless Pauli-Fierz action in Hamiltonian form is
\be\label{actionspin2res}
\bal
S[h_{ab}, \pi^{ab}, h_{00}, h_{0a}] &= \int d^D x\bigg( 
\pi^{ab} \dot h_{ab} - \mathcal H - \frac{1}{2c^2} h_{00} \mathcal{C} - h_{0}^{\; a}\mathcal{C}_a
\bigg),
\eal
\ee
 where $h_{ab}$ are the spatial components of the graviton field, $\pi^{ab}$ the conjugate momenta and $h_{00}$, $h_{0a}$ the Lagrange multipliers for the first-class constraints 
 \be
 \mathcal{C} =  - \partial^a \partial_a h_{b}^{\; b}+\partial^a \partial^b h_{ab} \approx 0, \qquad \mathcal{C}_a = - 2 \partial^b \pi_{ab} \approx 0.
\ee
We have rescaled $h_{00}$ as
\be
h_{00}\rightarrow c^2 h_{00},
\ee
so that $h_{00}$ and $h_{ab}$ have same dimension. 
Spatial indices are raised with $\delta^{ab}$. 

The energy density $\mathcal H$ reads
\be
\mathcal H = \mathcal{E}^{E} + \mathcal{E}^{M},
\ee
where the electric and magnetic contributions are respectively
\begin{eqnarray}
&& \mathcal{E}^{E} = c^2 \left(\pi^{ab} \pi_{ab}
-\frac{1}{D-2} (\pi^{a}_{\; a})^2\right) , \\
&&   \mathcal{E}^{M}= \frac{1}{4} \partial^a  h^{bc} \partial_a   h_{bc}  
- \frac12 \partial_b  h^{bc}\partial ^a  h_{ac}
+\frac12 \partial^a  h_{ab} \partial^b  h^{c}_{\; c}   
-\frac{1}{4} \partial^a  h^{b}_{\; b}  \partial_a  h^{c}_{\; c}.
\end{eqnarray}
The total energy $\mathcal{E}$ is not strictly invariant under the gauge transformations generated by the first-class constraints, which are,
\be
\delta \pi^{ab} = - \partial^a \partial^b \xi^0 + \delta^{ab} \partial^c \partial_c \xi^0 , \qquad \delta h_{ab} = \partial_a \xi_b + \partial_b \xi_a.
\ee
More precisely, $\mathcal{E}$ is only invariant up to the divergence $\partial_a V^a$ of a spatial vector and constraint terms,
\be
\delta \mathcal{E} =\partial_a V^a - c^2 \partial_a \xi^0 \mathcal{C}^a, \quad  V^a = - 2 c^2 \partial_b \xi^0 \pi^{ab} - \frac14 h^c_{\; c} \partial_b\partial^{[a} \xi^{b]} + h^{bc} \partial^a \partial_b \xi_c  - h^{ac} \partial^b \partial_b \xi_c.
\ee
Its integral over space is thus gauge invariant when the constraints hold. 

As in the case of $p$-forms, the electric energy density depends only on the momenta, while the magnetic energy density depends only on the fields.  Since the gauge transformation generated by $\mathcal{C}$ affects only the momenta $\pi^{ab}$, while those generated by $\mathcal{C}_a$ affects only the $h_{ab}$, one has separately that $\mathcal{E}^E$ and $\mathcal{E}^M$ are gauge invariant up to a divergence, so that their integrals over space are gauge invariant (on the constraint surface).

Since the constraints and hence the gauge invariances are unchanged in the limits, one can take consistently the magnetic and electric limits as in the previous section.  The magnetic limit amounts to setting $c = 0$ in the above expressions, while the electric limit needs a rescaling of the fields before setting $c$ equal to zero.

\subsubsection*{Mass term}

If we add the mass term $-\frac{m^2}{2}\left( \frac{1}{2} h_{ab}  h_{ab}-  \frac{1}{c^2} h_{0a}  h_{0a}+\frac{1}{c^2}  h_{00}  h_{aa} -\frac{1}{2} h_{aa}  h_{bb}  \right)$ one gets as new Hamiltonian action, after the $h_{0a}$'s, which are now auxiliary fields, are eliminated using their equations of motion,
\be\label{actionspin2resMass}
\bal
S_{m^2 \not= 0}[h_{ab}, \pi^{ab}, h_{00}] &= \int d^D x\bigg( 
\pi^{ab} \dot h_{ab} - \mathcal H' - \frac 12 h_{00} \mathcal{C}'
\bigg),
\eal
\ee
 with
 \be
\mathcal H' = \mathcal{E}'^{E} + \mathcal{E}'^{M},
\ee
and
\begin{eqnarray}
&& \mathcal{E}'^{E} = c^2 \pi^{ab} \pi_{ab}
-\frac{c^2}{D-2} (\pi^{a}_{\; a})^2 + \frac{2 c^2}{m^2} \big(\partial_a \pi^{ab} \partial^c \pi_{cb} \big), \\
&&   \mathcal{E}'^{M}= \frac{1}{4} \partial^a  h^{bc} \partial_a   h_{bc}  
- \frac12 \partial_b  h^{bc}\partial ^a  h_{ac}
+\frac12 \partial^a  h_{ab} \partial^b  h^{c}_{\; c}   
-\frac{1}{4} \partial^a  h^{b}_{\; b}  \partial_a  h^{c}_{\; c}\nonumber \\
&& \qquad \qquad \qquad + \frac{1}{4} m^2 \left(h^{ab}h_{ab} - (h^a_{\; a})^2\right),\\
&&\mathcal{C}' =  - \partial^a \partial_a h_{b}^{\; b}+\partial^a \partial^b h_{ab} + m^2 h^a_{\; a}.
\end{eqnarray}
There is now no gauge invariance since the equation of motion $\dot{\mathcal C}' = 0$ implies the constraint
$\mathcal D \approx 0$ with
\be
\mathcal D = \partial_a \partial_b \pi^{ab} + \frac{m^2}{D-2} \pi^a_{\; a}  .
\ee
The pair $(\mathcal{C}', \mathcal D)$ is second class.  

One can clearly take again consistently the electric and magnetic Carroll limits.

\subsection{Higher spins}

One can extend the procedure to Lorentz-invariant higher spin theories described by the Fronsdal action \cite{Fronsdal:1978rb}.  These possess also two Carroll contractions.  A new interesting feature emerges, however, which is that in order to preserve the form of the gauge invariances (i.e., of the constraints),  some of the original Lagrangian field components should be regarded as ``momenta'' (i.e., are $p$'s rather than $q$'s) while their conjugate variables should be regarded as  ``fields'' (i.e., are $q$'s rather than $p$'s).

We illustrate this phenomenon with the spin $3$ field.  Its Hamiltonian formulation reads, with the rescaling 
$\phi_{00a}\rightarrow c^2 \phi_{00a}$, 
\be\label{HAspin3}
S=\int d^D x\bigg(\Pi^{abc}\dot \phi_{abc} +\Pi \dot\alpha -\mathcal H
-\phi_{0ab}\mathcal{C}^{ab}
- \phi_{a00}\mathcal{C}^a \bigg),
\ee
where $\phi_{abc}$ are the spatial components of the spin-$3$ field,
\be
\alpha \equiv \frac{1}{c^2}\phi_{000} - 3 \phi_{0a}^{\; \; \; a},
\ee
and $\Pi^{abc}$, $\Pi$ their conjugate momenta.
Here, the energy density is explicitly given by (with $\phi_a \equiv \phi_{ak}^{\; \; \; k}$ and $\Pi^a \equiv \Pi^{ak}_{\; \; \; k}$)
\be
\bal
\mathcal H&=
c^4\Pi^2
+\frac{c^2}{2}\Pi_{abc} \Pi^{abc}
-\frac{3c^2}{2D}\Pi_{a}\Pi^a
+ \frac{5D-3}{8c^2D}\partial_{a}\alpha \partial^{a}\alpha
+\frac{3}{2D}\Pi^{a}\partial_a\alpha
\\&
+\frac{1}{2}\partial_k \phi_{abc} \partial^k \phi^{abc} 
-\frac{3}{2} \partial^a \phi^{bcd} \partial_b \phi_{acd}
+3\partial^a \phi_{abc} \partial^b \phi^{c}
-\frac{3}{2}\partial_a \phi_{b}\partial^a \phi^{b}
-\frac{3}{4}(\partial^a \phi_{a})^2,
\eal
\ee
and the constraints read
\be
\mathcal{C}_{ab} =- 3 \partial^c\Pi_{abc}
- \frac{3}{2c^2}\delta_{ab},
\Delta \alpha
\ee
and
\be
\mathcal{C}_a= \
3c^2\partial_a\Pi 
+3\partial^b\partial^c  \phi_{abc} 
-3\Delta \phi_{a}
-\frac{3}{2}\partial_a \partial^b \phi_{b}.
\ee
The temporal components $\phi_{0bc}$ and $\phi_{00c}$ are the Lagrange multipliers for the constraints.  These are first class and generate the following gauge transformations:
\begin{itemize}
\item Constraint $\mathcal{C}_{ab}$:
\begin{eqnarray}
&& \delta_\epsilon h_{ijk} = 3 \partial_{(i} \epsilon_{jk)}, \\
&& \delta_\epsilon \Pi = \frac{1}{c^2}\frac32 \triangle \epsilon^a_{\; a}
\end{eqnarray}
($\delta_\epsilon \pi^{ijk} = 0$, $\delta_\epsilon \alpha = 0$).
\item  Constraint $\mathcal{C}_a$:
\begin{eqnarray}
\hspace{-.4cm} &&\delta_\chi \Pi^{ijk} = -  \partial^{(i} \partial^j \chi^{k)} +  \delta^{(ij}  \left(  \triangle \chi^{k)} + \frac12 \partial^{k)} \partial_m \chi^m \right)\hspace{.7cm}\\
\hspace{-.4cm}&& \delta_\chi \alpha = - c^2 \partial_m \chi^m  
\end{eqnarray}
($\delta_\chi h_{ijk} = 0$, $\delta_\chi \Pi = 0$).
\end{itemize}

The fields ($h_{ijk}, \Pi)$ transform together, and only under the action of $\mathcal{C}_{ab}$; the fields ($\pi^{ijk}, \alpha)$ transform together, and only under the action of $\mathcal{C}_{a}$.  However, $c^2$ appears in the transformation laws.  In order to avoid a singular limit as $c \rightarrow 0$ and, more crucially, a modification of the constraints and the gauge transformations in the Carrollian limit, we rescale $\alpha$ and  its conjugate momentum (besides the already performed rescaling of the Lagrange multiplier $\phi_{a00}$)
\be
\alpha\rightarrow c^2 \alpha , 
\qquad
\Pi\rightarrow \frac{1}{c^2}\Pi  , 
\ee
which has the effect of exchanging the role of $\alpha$ and its conjugate momentum $\Pi$. This yields
\be\label{HAspin3resc}
S=\int d^D x\bigg(\Pi_{abc}\dot \phi_{abc} +\Pi \dot\alpha -\mathcal H
-\phi_{0ab}\mathcal{C}^{ab}
-\frac{1}{c^2}\phi_{a00}\mathcal{C}^a \bigg),
\ee
where the constraints are now $c$-independent, 
\be
\mathcal{C}_{ab} =- 3 \partial^c\Pi_{abc}
- \frac{3}{2}\delta_{ab}
\Delta \alpha,
\ee
and
\be
\mathcal{C}_a= \
3\partial_a\Pi 
+3\partial^b\partial^c  \phi_{abc} 
-3\Delta \phi_{a}
-\frac{3}{2}\partial_a \partial^b \phi_{b}.
\ee
The energy density reads
\be\label{Hspin3resc}
\mathcal H= \mathcal{E}^E + \mathcal{E}^E,
\ee
with
\be
\mathcal E^E=
c^2 \left(\frac{1}{2}\Pi_{abc} \Pi^{abc}
-\frac{3}{2D}\Pi_{a}\Pi^a
+ \frac{5D-3}{8D}\partial_{a}\alpha \partial^{a}\alpha
+\frac{3}{2D}\Pi^{a}\partial_a\alpha \right),
\ee
and
\be
\mathcal E^M=
\Pi^2
+\frac{1}{2}\partial_k \phi_{abc} \partial^k \phi^{abc} 
-\frac{3}{2} \partial^a \phi^{bcd} \partial_b \phi_{acd}
+3\partial^a \phi_{abc} \partial^b \phi^{c}
-\frac{3}{2}\partial_a \phi_{b}\partial^a \phi^{b}
-\frac{3}{4}(\partial^a \phi_{a})^2.
\ee

One can easily take the Carrollian limits in a way that manifestly preserves the gauge symmetries.  The magnetic limit is obtained by dropping $\mathcal E^E$ while the electric limit is obtained by dropping $\mathcal E^M$, after rescalings analogous to those of the $p$-form case.  Even though we swapped the roles of $\alpha$ and its conjugate momentum $\Pi$ before taking the contractions, Carroll causality is easily verified to hold in the limit because the dynamical equations effectively reduce again to ordinary differential evolution equations with respect to time.

\section{Manifestly Carroll invariant actions for $p$-form gauge fields}
\label{sec:manifestCov}

The Hamiltonian actions considered so far do not exhibit explicitly spacetime covariance in the sense that we cannot use directly standard tensor calculus to check their Carroll invariance. A manifestly spacetime covariant action can be useful in certain circumstances.  We achieve the task of constructing manifestly Carroll invariant actions for $p$-form gauge fields, dealing first with the cases $p=0$ and $p=1$ and generalizing then to all $p$'s. 

\subsection{Scalar field}

\subsubsection{Electric contraction}

We start with the simpler electric case.

In the electric limit, the manifestly covariant action is just obtained by eliminating the momentum $\pi_\phi$ in terms of the velocity $\dot{\phi}$ using its own equation of motion.  One gets $S^E[\phi] = \frac12 \int d^D x \dot{\phi}^2$, or in covariant form
\be
S^E[\phi] = \frac12 \int d^D x (n^\alpha \partial_\alpha \phi)^2
\ee
This action is manifestly invariant under Carroll transformations, which leave $n^\alpha$ invariant and preserve the volume element.

The field $\phi$ is a scalar, the components $\partial_\alpha \phi$ are the components of a $1$-form and the momentum $\pi_\phi = n^\alpha \partial_\alpha \phi$ is a scalar.

\subsubsection{Magnetic contraction}
In the magnetic limit, one cannot eliminate $\pi_\phi$ using its own equation of motion.  In fact, the connection between $\dot{\phi}$ and $\pi_\phi$ is lost, since $\dot{\phi} =0$ but in general $\pi_\phi \not=0$.

Therefore, we try to rewrite the first-order action directly in manifestly covariant form.    To that end, we recall that if a covariant tensor is transverse (its contraction with $n^\alpha$ on any index is zero), then its square norm is well defined.  Thus, at least on shell, where $n^\alpha \partial_\alpha \phi = 0$, one can rewrite the energy density of the magnetic theory as $\partial_\alpha \phi  \partial_\beta \phi G^{\alpha \beta}$, where $G^{\alpha \beta}$ is any contravariant tensor fulfilling (\ref{eq:DefG0}), i.e., 
$
G^{\alpha \beta} g_{\beta \gamma} = \delta^\alpha_\gamma - n^\alpha \theta_\gamma
$.
We need, however, to define the action off-shell.  For that purpose, we introduce the one-form $\theta_\alpha(x)$ (with $\theta_\alpha n^\alpha = 1$), which we treat as a dynamical variable and consider the action
\be
S^M[\phi, \pi_\phi, \theta_\alpha] = \int d^Dx \big(\pi_\phi n^\alpha \partial_\alpha \phi - \frac12 G^{\alpha \beta} \partial_\alpha \phi \partial_\beta \phi \big). \label{eq:MagnAction1}
\ee
This action is Carroll invariant if we transform $\phi$, $\pi_\phi$ as scalars and $\theta_\alpha$ as a one-form.

It is nevertheless not clear that it is satisfactory since it involves the extra field $\theta_\alpha$, which might change the dynamics.  It turns out, however, that this is not the case and that the action (\ref{eq:MagnAction1}) is dynamically equivalent to the Hamiltonian action (\ref{eq:MagnAction0}).   This is because it possesses a gauge invariance that enables one to gauge the extra field $\theta_\alpha$ away,  thereby reducing (\ref{eq:MagnAction1}) to (\ref{eq:MagnAction0}).  Indeed, if we shift $\theta_\alpha$ as in (\ref{eq:TransfGInf}) (with $\lambda_\alpha n^\alpha = 0$), the term $\frac12 G^{\alpha \beta} \partial_\alpha \phi \partial_\beta \phi $ changes as
\be
\delta \theta_\alpha = \lambda_\alpha, \qquad \delta \left(\frac12 G^{\alpha \beta} \partial_\alpha \phi \partial_\beta \phi \right) = -n^\alpha G^{\beta \rho} \lambda_\rho \partial_\alpha \phi \partial_\beta \phi
\ee
If we transform at the same time the momentum $\pi_\phi$ as
\be
\delta \pi_\phi = - G^{\beta \rho} \lambda_\rho \partial_\beta \phi
\ee
the action is invariant.  

Using this gauge invariance, we can set $(\theta_\alpha) = (1, 0, \cdots, 0)$, in which case $G^{\alpha \beta}$ takes the canonical form (\ref{eq:FlatCarrollG}) and the action (\ref{eq:MagnAction1}) reduces to (\ref{eq:MagnAction0}).  Note that the gauge condition $\theta_a = 0$ is not Carroll invariant.  Under a Carroll boost parametrized by the vector field $\xi = b_a x^a \frac{\partial}{\partial x^0}$, one finds $\delta \theta_\alpha = \mathcal{L}_\xi \theta_\alpha = (0,  b_a)$, and thus one must accompany the Carroll boost by the compensating gauge transformation with $\lambda_a =- b_a$ to maintain $\theta_a = 0$.  This means that once the gauge is fixed, the momentum $\pi_\phi$ transforms as
\be\label{eq:deltapi}
\delta_b \pi_\phi =  \mathcal{L}_\xi \pi_\phi + b^\alpha \partial_\alpha \phi = b_a x^a \partial_0 \pi_\phi + b^\alpha \partial_\alpha \phi\qquad b^\alpha = (0, b^a) 
\ee
where $\mathcal{L}_\xi \pi_\phi$ is here the Lie derivative of $\pi_\phi$ viewed as a scalar (ordinary transport term $\xi^\rho \partial_\rho \pi_\phi$) and where the second term comes from the compensating gauge transformation.

This is a perfectly acceptable transformation rule.  In fact, if we construct the $D$-component object with components $(V^0 = \pi_\phi, V^a = - \partial^a \phi)$ (index raised with $\delta^{ab}$), the transformation law  (\ref{eq:deltapi}) coincides with the zeroth-component of $\mathcal{L}_\xi V^\alpha$ where $V^\alpha$ are viewed as the components of a contravariant $D$-vector, i.e., $\delta_b \pi_\phi  = \mathcal{L}_\xi V^0= \xi^\rho \partial_\rho V^0 - V^\rho \partial_\rho \xi^0$ -- and the spatial components of the relation $\delta_b V^a =  \mathcal{L}_\xi V^a$ are obviously fulfilled since  $\xi^a = 0$.  Thus, the procedure of gauge-fixing has effectively changed the transformation law of $\pi_\phi$ from that of a scalar to that of the zeroth-component of the contravariant vector field $V^\alpha$.

If we compare the transformation laws of $\pi_\phi$ in the electric and magnetic theories, we see that they are different.  In the electric case, $\pi_\phi$ transforms as the zeroth component of a covariant vector field, while in the magnetic case, it transforms as the zeroth-component of a contravariant vector field.  The two representations are inequivalent and there is indeed no non-degenerate invariant metric to go from one to the other.
The same difference in transformation rules was observed and analyzed in the case of electromagnetism, where again, the electric theory was naturally found to correspond to a formulation involving fields transforming in covariant representations, while the magnetic theory was found to correspond to contravariant fields  \cite{Duval:2014uoa}.  How this arises in our approach will be discussed in the next section, but first, we show that the same conclusions concerning transformation rules hold in the Hamiltonian formalism.

One way to characterize the inequivalence of the two representations is to observe that in the representation described by covariant vectors , there is a $d$-dimensional invariant subspace defined by $V_0=0$.  The representation is not completely reducible since there is no complementary invariant one-dimensional subspace (the conditions $V_a = 0$ are not invariant), i.e., is not decomposable.  In the dual contravariant case, there is an invariant one-dimensional subspace defined by $V^a=0$, but no invariant complementary $d$-dimensional subspace; the representation is again indecomposable, but in a different way.

What makes the above construction possible, is that the representation in the $d$-dimensional invariant subspace of the covariant representation is equivalent to the $d$-dimensional quotient representation of the contravariant representation by its one-dimensional invariant subrepresentation.  This $d$-dimensional representation represents trivially the boosts and coincides with the vector representation of the spatial rotation group, for which there is no distinction between covariant and contravariant tensors.

To conclude this Section,  we show equivalence of the transformation rules derived in the covariant formulation and in the Hamiltonian formulation.  We only need to consider Carroll boosts since the other Carroll transformations raise no particular difficulty.

Under a Carroll boost parametrized by the vector field $\xi=  b_k x^k \frac{\partial }{\partial x^0} \Leftrightarrow \xi^0 = b_k x^k, \xi^k = 0$,  the Hamiltonian fields transform as $\delta \phi^A = [\phi^A, \int d^dx b_k x^k \mathcal{E} ]$.   This must be compared with  $\delta \phi^A = \mathcal{L}_\xi \phi^A$.

In the electric case where $\pi$ is the zeroth component $\partial_0 \phi$ of the one-form $\partial_\alpha \phi$, one gets
\be
\delta_b \phi = b_k x^k \pi_\phi , \qquad \delta_b \pi_\phi = 0,
\ee
which is found to be in perfect agreement with $\delta_b \phi = \mathcal{L}_\xi  \phi= \xi^\rho \partial_\rho \phi$  and $\delta_b \partial_\alpha \phi = \mathcal{L}_\xi \partial_\alpha \phi = \xi^\rho \partial_\rho (\partial_\alpha \phi) + \partial_\alpha \xi^\rho \partial_\rho \phi$ if one uses the equations of motion.

In the magnetic case, one gets
\be
\delta_b \phi = 0, \qquad \delta_b \pi_\phi =b^k\partial_k\phi+ b_j x^j \partial_k\partial^k\phi
\ee
Again, this is in perfect agreement with $\delta_b \phi = \mathcal{L}_\xi  \phi$ and $\delta_b V^\alpha = \mathcal{L}_\xi V^\alpha= \xi^\rho \partial_\rho V^\alpha - V^\rho \partial_\rho \xi^\alpha$ with $V^\alpha = (\pi_\phi, -\partial^a \phi)$ when one uses the equations of motion.

 \subsection{Electromagnetism}
 
 \subsubsection{Electric contraction}
 
 The actions of the electromagnetic contractions can also be cast  in  a manifestly Carroll-covariant form.  Again, we start with the simpler electric limit, where 
one can eliminate the momenta by means of their own field equations. If one does this, one finds that the action, which is second order,  takes the form
\be
S^E[A_i, A_0]=\frac{1}{2}\int d^D x F_{0i}F_0^{\;\;i}
\ee
or, in manifestly Carroll-invariant form,
 \be
 S^E[A_\alpha] = \frac12 \int d^D x (n^\alpha F_{\alpha \beta}) ^2
 \ee
 The integrand $G^{\rho \sigma}  n^\alpha F_{\alpha \rho} n^\beta F_{\beta \sigma}$ is well-defined because $F_{\alpha \beta}n^\beta$ is transverse, $n^\alpha F_{\alpha \beta} n^\beta = 0$.
 
The electric and magnetic fields form the components of the covariant antisymmetric tensor $F_{\alpha \beta}$, as in the Maxwell theory.  The action is invariant under 
 \be
 \delta A_\alpha = \mathcal{L}_\xi A_\alpha = \xi^\rho \partial_\rho A_\alpha + \partial_\alpha \xi^\rho A_\rho
 \ee
 which implies
\be
\delta F_{\alpha \beta} = \mathcal{L}_\xi F_{\alpha \beta} = \xi^\rho \partial_\rho F_{\alpha \beta}+ \partial_\alpha \xi^\rho F_{\rho \beta} + \partial_\beta \xi^\rho F_{\alpha \rho} 
\ee
 In particular for Carroll boosts $\xi= b_a x^a \frac{\partial}{\partial x^0}$, one gets
 \be
 \delta A_0 = b_a x^a \dot{A}_0, \qquad \delta A_k = b_a x^a \dot{A}_k + b_k A_0
 \ee
 which leads  to the transformations of the electric and magnetic fields
 \be
 \delta F_{0k} =  b_a x^a \dot{F}_{0k}, \qquad \delta F_{km} =  b_a x^a \dot{F}_{km}+ b_k F_{0m} - b_m F_{0k}
 \ee
 equivalent to the expressions given in \cite{Duval:2014uoa}.

 \subsubsection{Magnetic  contraction}
 
 The magnetic Carroll limit of electromagnetism is very similar to the magnetic Carroll limit of the Klein-Gordon theory, with in particular, the impossibility to express the momenta $\pi^i$ in terms of the velocities through their equations of motion.
 
 We thus follow the same steps as in the scalar case, and look for a direct covariantization of the first-order Hamiltonian action.  For that purpose, we introduce the gauge field $\theta_\alpha$ that enables one to define $G^{\alpha \beta}$.  We also assume initially that the momenta $\pi^a$ are the spatial components of a spacetime vector $\pi^\alpha$, with the gauge invariance $\pi^\alpha \rightarrow \pi^\alpha + \lambda n^\alpha$ ($\lambda$ arbitrary) to keep the number of degrees of freedom unchanged.  As we shall see, a representation transmutation phenomenon similar to the one found in the scalar case will occur for $\pi^\alpha$.  
 
We postulate the action
 \be
 S^M[A_\alpha, \pi^\beta, \theta_\gamma] =  \int d^D x \left( \pi^\alpha F_{\alpha \beta} n^\beta - \frac14 G^{\alpha \beta} G^{\rho \sigma} F_{\alpha \rho} F_{\beta \sigma}\right) \label{eq:MagnAction2}
 \ee
The gauge invariance $\pi^\alpha \rightarrow \pi^\alpha + \lambda n^\alpha$ is obvious since $F_{\alpha \beta}$ is antisymmetric - $\pi^0$ just drops.  

The scalar product $G^{\alpha \beta} G^{\rho \sigma} F_{\alpha \rho} F_{\beta \sigma}$ is $\theta$-independent when the equations of motion for $\pi_\alpha$ hold, but off-shell, this scalar product -- and hence also the action -- does depend on $\theta_\alpha$. 
This dependence, however, is associated with a gauge invariance, just as in the scalar case.  If one shifts $\theta_\alpha$ as in (\ref{eq:TransfGInf}) (with $\lambda_\alpha n^\alpha = 0$) and at the same time transforms $\pi^\alpha$ as
\be
\delta \theta_\alpha = \lambda_\alpha, \qquad \delta \pi^\alpha = - G^{\alpha \rho} F_{\rho \sigma} \lambda^\sigma
\ee
the action is invariant.

Therefore, we can shift away $\theta_\alpha$ and go to the gauge $\theta_0 = 1, \theta_a= 0$.  In that gauge, the action ( \ref{eq:MagnAction2}) reduces to the Carroll magnetic action (\ref{eq:MagneticAction00}). Furthermore, just as in the scalar case, the gauge condition is not maintained by Carroll boosts, which must be supplemented by a compensating $\theta$-shift to bring one back to $\theta_0 = 1, \theta_a= 0$. Under this compensating gauge transformation, $\pi^a$ picks up a term of the form $ F^{ac} b_c$.   The net result is that the electric field and the magnetic field transform in the representation of the Carroll group given by antisymmetric contravariant tensors $H^{\alpha \beta}$, in agreement with \cite{Duval:2014uoa} (note the obvious typo in (5.14) of that reference, the transformation of the electric field should involve B instead of E in the second term).  One has $(H^{\alpha \beta}) = (\pi^a, F^{ab})$ and
 \be
 \delta H_{0k} =  b_a x^a \dot{H}^{0k} - b_a H^{ak}, \qquad \delta H_{km} =  b_a x^a \dot{H}^{km}
 \ee
under Carroll boosts (the first term is the standard transport term, the second term is determined by the representation). 

One can also verify that the Carroll transformations are correctly generated in the Hamiltonian formalism. Note that as it is well known, the transformation of the vector potential differs from its Lie derivative by a gauge transformation that drops when computing the transformation of gauge invariant quantities.

 \subsection{$p$-form gauge theories}
 \label{sec:pForms}

 Direct extension of the previous derivations yield as covariant action
 \be
 S^E[A_{\alpha_1 \cdots \alpha_p}] = \frac{1}{2 p!} \int d^D x (n^\beta F_{\beta \alpha_1 \cdots \alpha_p}) ^2
 \ee
 for the electric-type contraction and  
  \be
 S^M[A_{\alpha_1 \cdots \alpha_p}, \pi^{\beta_1 \cdots \beta_p}, \theta_\gamma] =  \int d^D x \left( \pi^{\alpha_1 \cdots \alpha_p} n^\beta F_{  \beta \alpha_1 \cdots \alpha_p }  - \frac{1}{2 (p+1)!}( F_{\alpha_1 \cdots \alpha_p \alpha_{p+1}} )^2\right) \label{eq:MagnActionpForm}
 \ee
 for the magnetic-type one.  Here, in computing expressions such as $(n^\beta F_{\beta \alpha_1 \cdots \alpha_p}) ^2$ or $( F_{\alpha_1 \cdots \alpha_p \alpha_{p+1}} )^2$, one raises of course  the indices with $G^{\alpha \beta}$.
The form of the gauge transformation that shifts $\theta_\alpha$ takes now the form
\be
\delta \theta_\alpha = \lambda_\alpha, \qquad \delta \pi^{\alpha_1\cdots\alpha_p} = -\frac{1}{p!} G^{\alpha_1 \beta_1}\cdots G^{\alpha_p \beta_p}  F_{\beta_1 \cdots \beta_p \sigma} \lambda^\sigma \, .
\ee

  \section{Carroll contractions of Einstein's theory}
\label{sec:Gravity}

The Carroll contractions of gravity are discussed along similar lines once the Einstein action is put in Hamiltonian form. Spacetime covariance corresponds now to a local symmetry generated by first-class constraints.  The question from the Hamiltonian viewpoint, then, is to check whether there are contractions of these constraints that yield the Carroll structure.  It is easy to show that this is so.

The Hamiltonian action reads, in standard notations,
\be
S[g_{ij}, \pi^{ij}, N, N^i] = \int dx^0 \int d^d x (\pi^{ij} \dot{g}_{ij}  - N  \mathcal{H} - N^i \mathcal{H}_i)
\ee
 where we do not write explicitly the surface terms as these will be discussed elsewhere when we analyse the asymptotic symmetries. Here, $\mathcal{H}\approx 0$ is the Hamiltonian constraint and $\mathcal{H}_i \approx 0$ is the momentum constraint.  The explicit expressions are, in appropriate units,
 \be 
 \mathcal{H} = G_{ijkm} \pi^{ij} \pi^{mn} - R \sqrt{g}, \qquad \mathcal{H}_i  = -2 \pi_{i \; \;  \vert j}^{\; j}.
 \ee

 One can drop consistently either term in the Hamiltonian constraint, since in each case, one gets the system of first class constraints,
 \begin{eqnarray}
 && [\mathcal{H}^{C}(x), \mathcal{H}^{C}(x')] = 0 , \\
 &&  [\mathcal{H}^C(x), \mathcal{H}_k(x')] = (\mathcal{H}^C(x) + \mathcal{H}^C(x')) \delta_{,k}(x- x') \\
 && [\mathcal{H}_m(x), \mathcal{H}_k(x')] = \mathcal{H}_m(x') \delta_{,k}(x- x')+ \mathcal{H}_k(x) \delta_{,m}(x- x')
 \end{eqnarray}
 where $\mathcal{H}^C$ stands for either $\mathcal{H}^E$ or $\mathcal{H}^M$,
 \be
 \mathcal{H}^E = G_{ijkm} \pi^{ij} \pi^{mn}, \qquad  \mathcal{H}^M = - R \sqrt{g}
 \ee
 This first class constraint algebra is precisely the algebra characteristic of Carrollian spacetimes (see \cite{Teitelboim:1972vw},\cite{Teitelboim:1978wv,Henneaux:1979vn}) and therefore, there are again two consistent Carroll contractions.  Note that the cosmological constant term $\Lambda \sqrt{g}$ is consistently allowed in both limits.  The electric-type contraction, where spatial gradients are dropped in $\mathcal{H}$, is the strong coupling limit \cite{Isham:1975ur}, or zero signature limit \cite{Teitelboim:1978wv} defined long ago, which are relevant to the BKL behaviour.

 The Hamiltonian action possesses in each case  $D$ class constraints, which correctly matches the number of gauge invariances of a diffeomorphism invariant theory. The manifestly covariant action  (i.e., in the present case, manifestly diffeomorphism invariant action) for the electric limit was written in \cite{Henneaux:1979vn}.  It involves the second fundamental form $K_{\alpha \beta}$ defined as $(- \frac12)$ times the Lie derivative of the degenerate metric $g_{\alpha \beta}$ along the vector $n^\alpha$,
 \be
 K_{\alpha \beta} = - \frac12 \mathcal{L}_n g_{\alpha \beta}
 \ee
 and reads
 \be
 S^{E}[g_{\alpha \beta}, \Omega] = \int d^Dx \Omega (K^{\alpha \beta} K_{\alpha \beta} - K^2)
 \ee
 This expression makes sense because $K_{\alpha \beta}$ is identically transverse, $K_{\alpha \beta} n^\beta = 0$, so that $K \equiv K_{\alpha \beta} G^{\alpha \beta}$ and $K^{\alpha \beta}K_{\alpha \beta} = K_{\rho \sigma} G^{\rho \alpha } G^{\sigma \beta}K_{\alpha \beta}$ are well defined.
 
 We have not derived the manifestly covariant action for the magnetic limit, where one gets the equation of motion $K_{\alpha \beta} = 0$ so that there is no connection between the momenta and the time derivatives of the metric.  
 
 In fact, covariant actions for Carroll gravity have been constructed in \cite{Hartong:2015xda,Bergshoeff:2017btm} by gauging the Carroll algebra\footnote{Covariant actions for Carroll gravity in three and two spacetime dimensions in have also been constructed in \cite{Bergshoeff:2016soe,Ravera:2019ize,Gomis:2019nih,Grumiller:2020elf,Gomis:2020wxp,Concha:2021jnn}.}.  Although we have not performed the explicit check, we suspect that the action of \cite{Hartong:2015xda}, which has a structure similar to $K^{\alpha \beta} K_{\alpha \beta} - K^2$, is equivalent to the above electric action, while the action of \cite{Bergshoeff:2017btm}, which implies $K_{\alpha \beta} = 0$, would be equivalent to the Hamiltonian action of the magnetic-type contraction. We hope to return to this issue in the future.

\section{Conclusions}
\label{sec:Conclusions}

In this paper, we have shown that Lorentz-invariant theories possess two distinct Carroll limits, one electric and one magnetic.  This generalizes to arbitrary fields what was found earlier for electromagnetism \cite{Duval:2014uoa}, and is the analog of a similar phenomenon described in the Galilean case \cite{LeBellac1973}.  The existence of two distinct limits reflects the fact that while contravariant and covariant tensors transform in equivalent representations of the Lorentz group where there is an invertible invariant metric connecting the two, this property no longer holds in the Carrollian case \cite{Duval:2014uoa}. 

Our approach for taking the Carrollian limits is based on the variational formulation and provides automatically Carroll invariant action principles.   Spacetime covariance is not manifest, however, since we use the Hamiltonian form of the variational principle.  It is controlled through the Poisson brackets of the energy density and momentum density.  We have nevertheless constructed covariant action principles for $p$-form gauge theories for both the electric magnetic limits, which share, in the electric case, features quite similar to those of the electric-type limit of Einstein theory \cite{Henneaux:1979vn}. 

As we alluded to above, a different method for constructing Carroll-invariant actions have been devised more recently for Carrollian gravities, by gauging the Carroll algebra \cite{Hartong:2015xda,Bergshoeff:2017btm}. It would be interesting to explicitly compare  this approach with our results, as well as its extension to the higher spin Carroll algebras of \cite{Campoleoni2021}.  

Since one potentially useful application of the Carroll algebra deals with non relativistic holography, another interesting problem is to perform the asymptotic analysis at spatial infinity of both the electric and magnetic limits of Einstein gravity in arbitrary spacetime dimensions.

\section*{Acknowledgements}

We thank Andrea Campoleoni for useful discussions  on higher spin Carroll algebras and Joaquim Gomis for interesting comments on Carroll particles.  We are also grateful to the Erwin Schr\"odinger International Institute for Mathematics and Physics (ESI) for kind hospitality which this work was being completed. This research has been partially supported by the ERC Advanced Grant ``High-Spin-Grav'', by FNRS-Belgium (conventions FRFC PDRT.1025.14 and IISN 4.4503.15), as well as by funds from the Solvay Family.

\bibliographystyle{elsarticle-num}

\bibliography{\jobname}

\end{document}